\documentclass[12pt,preprint]{aastex}
\def\oh{(O/H)$_{\rm{gas}}$}

\newcommand{\lya}{Lyman-$\alpha$}
\newcommand{\htwo}{H$_2$}
\slugcomment{Version \today}
\shorttitle{Oxygen Gas Phase Abundance Revisited}
\shortauthors{Andr\'e et al.}

\begin{document}
\title{Oxygen Gas Phase Abundance Revisited}
\author{M.~K. Andr\'e\altaffilmark{1,2},~C.~M. Oliveira\altaffilmark{2}}
\author{J.~C. Howk\altaffilmark{2},~R. Ferlet\altaffilmark{1},~J.--M. D\'esert\altaffilmark{1},
~G. H\'ebrard\altaffilmark{1}}
\author{ S. Lacour\altaffilmark{2},~A. Lecavelier des \'Etangs\altaffilmark{1},~A. Vidal-Madjar\altaffilmark{1},
~H.~W. Moos\altaffilmark{2}}

\altaffiltext{1}{Institut d'Astrophysique de Paris, 98 bis bd Arago, 75014 Paris, France, andre@iap.fr}
\altaffiltext{2}{The Johns Hopkins University, 3400 N. Charles Street, Baltimore, MD 21218, USA}

\begin{abstract}
We present new measurements of the interstellar gas-phase oxygen abundance
along the sight lines towards 19 early-type galactic stars at
an average distance of 2.6 kpc. We derive O~{\small I} column densities
from {\it HST}/STIS observations of the weak 1355 \AA~ intersystem
transition. We derive total hydrogen column densities [N(H~{\small I})+2N(H$_2$)]
using {\it HST}/STIS observations of \lya~ and {\it FUSE}
observations of molecular hydrogen. The molecular hydrogen content of
these sight lines ranges from f(H$_2$) = 2N(H$_2$)/[N(H~{\small I})+2N(H$_2$)] = 0.03 to 0.47.
The average $\langle H_{tot}/E_{B-V} \rangle$ of 6.3$\times$10$^{21}$ cm$^{-2}$
mag$^{-1}$ with a standard deviation of 15\% is consistent with previous surveys.
The mean oxygen abundance along these sight lines, which probe a wide range of galactic
environments in the distant ISM, is 10$^6$ \oh~= $408~\pm~13$ (1~$\sigma$ in the mean).
We see no evidence for decreasing gas-phase oxygen abundance with increasing
molecular hydrogen fraction and the relative constancy of \oh~suggests that
the component of dust containing the oxygen is not readily destroyed. We estimate that,
if 60\% of the dust grains are resilient against destruction by shocks, the distant interstellar total oxygen abundance 
can be reconciliated with the solar value derived from the most recent measurements
of 10$^6$ \oh$_\odot$ = 517~$\pm$~58 (1~$\sigma$). We note that the smaller oxygen abundances derived for the 
interstellar gas within 500 pc
or from nearby B star surveys are consistent with a local elemental
deficit.
\end{abstract}



\section{Introduction}
The chemical evolution of galaxies is a long-standing open question in astronomy (Audouze \& Tinsley 1976).
Of all atomic species, oxygen, which is mainly produced in type II supernovae,
has proven to be the most useful element for probing the astration of the ISM by massive
stars, both because of its large abundance and origin (for a review see Henry \& Worthey 1999). The value of the oxygen
abundance within a few hundred parsec from the Sun has been thoroughly investigated in the past two
decades toward dozens of sight lines with {\it Copernicus} (De Boer et al. 1981, York et al. 1983, Keenan et al. 1985),
{\it HST} (Meyer et al. 1998, Cartledge et al. 2001) and {\it FUSE} (Moos et al. 2002, and
references therein).
The accurate study by Meyer et al. (1998) revealed a well mixed medium
within 500 pc from the Sun : $\langle {\rm O/H} \rangle = 343\pm~15$ ppm (error in the mean). Note that this
value has been corrected linearly with an updated f-value (see \S 3.3). However, investigations
of more distant sight lines with the FUV absorption technique were limited by the unknown higher
molecular fractions and the fainter background continua.

Measurements of oxygen abundances beyond 1 kpc have been carried out
in two different environments : H~{\small II} regions and B star atmospheres. To this day,
these methods do not agree on a single picture for the oxygen distribution in the disk.
In the most recent survey of the oxygen abundance in galactic H~{\small II} regions, Deharveng et al.
(2001) confirm the existence of a metalicity gradient from the center out to 18 kpc in the outer
disk and find a local abundance consistent with ISM investigations. However they
point out that their value is highly dependent on the model adopted for the electron temperature
in the H~{\small II} region. As for the B star surveys, the considerable scatter in the measurements makes it
difficult to pin down an accurate value for the gradient (Smartt \& Rolleston 1997, Primas et al. 2001) while, at
the same time, the origin of the scatter is still debated.

With the high sensitivity of {\it FUSE} (10,000 times more sensitive than {\it Copernicus}; see Moos et al. 2000)
it is now possible to probe the molecular hydrogen content of the ISM beyond 1 kpc and in denser media as well.
Along with \lya~data from the STIS instrument, this allows us to obtain accurate total hydrogen column densities and 
hence derive a reliable {\rm O/H} ratio. Since oxygen and hydrogen have about the same ionization potential,
no ionization correction is needed and the analysis is largely model independent.
The principal question we try to answer in this work is the robustness of a constant {\rm O/H} for
large sight line distances.
The sight lines studied sample a vast diversity of environments. The enhanced depletion of oxygen in
dust grains along the densest
sight lines, the imprint of the galactic gradient along the most distant sight lines and
inefficient mixing processes are expected to play significant roles in the final scatter.

In this paper, we present new {\rm O/H} measurements toward 19 lines of sight (Table 1) obtained with data
from {\it FUSE} and STIS. For each one of these sight lines we performed a complete analysis of the O~{\small I},
H~{\small I}, and H$_2$ content, using different techniques.
The datasets and the reduction processing are described in \S 2. The analysis of O~{\small I}, H~{\small I}, and
H$_{2}$ are presented in \S 3. In \S 4 the results are discussed in view of the most recent investigations. Discussions
about the dust content and local oxygen deficit are given in \S 5 and conclusions are given in \S 6.

\section{Observations}

\subsection{STIS observations}

In this work, we use high-resolution STIS echelle observations of 19
stars (Table 1) to derive the column densities of H~{\small I}, O~{\small }I, and other atomic
species. The design and construction of STIS are described by
Woodgate et al. (1998), while information about the on-orbit
performance of STIS is summarized by Kimble et al. (1998). A summary of
the STIS observations is given in Table 2.

All observations in this work used the far-ultraviolet MAMA detector with the E140H
grating. Because the datasets are drawn from several programs, three
different apertures were used in obtaining the data. The
$0\farcs1\times0\farcs03$ aperture (the so-called ``Jenkins slit'')
provides a spectral resolution of $R\equiv \lambda/\Delta \lambda
\approx 200,000$ (velocity resolution of $\Delta v \sim
1.5$~km~s$^{-1}$, FWHM). These datasets have been discussed by
Jenkins \& Tripp (2001). The remainder of the data, taken through the
$0\farcs2\times0\farcs09$ or $0\farcs2\times0\farcs2$ apertures,
provide a spectral resolution of $R \approx 110,000$ (velocity
resolution of $\Delta v \sim 2.7$~km~s$^{-1}$, FWHM). Data taken
through the latter apertures have a line spread function (LSF) with
power in a halo extending $\sim \pm~5$ km~s$^{-1}$ from
line center. The LSF for all of the data have weak, broad wings stretching to
$\pm~10$ km~s$^{-1}$ (see the STIS Instrument Handbook).

The data were calibrated and extracted and the background estimated as
described by Howk \& Sembach (2000a). For spectral regions covered in
multiple orders, or when more than one observation existed, we
co-added the flux-calibrated data weighting each spectrum by the
inverse square of its error.

\subsection{{\it FUSE} observations}

This work makes use of {\it FUSE} observations for deriving molecular
hydrogen column densities along the 19 lines of sight studied. These
column densities are in turn used to derive the total (atomic+molecular)
hydrogen column densities along each sight line.  

The {\it FUSE} mission, its planning and on-orbit performance are discussed
by Moos et al. (2000) and Sahnow et al. (2000). Briefly, the {\it FUSE}
observatory consists of four co-aligned prime-focus telescopes and
Rowland-circle spectrographs that produce spectra over the wavelength
range 905-1187~\AA~with a spectral resolution of $\sim$~15-20
km~s$^{-1}$ for point sources (depending on the wavelength). Two of
the optical channels employ SiC coatings providing reflectivity in the
wavelength range $\sim$~905-1090~\AA~while the other two have LiF
coatings for maximum sensitivity above 1000~\AA. Dispersed light is
focused onto two photon-counting microchannel plate detectors.

Table 3 presents a summary of the {\it FUSE} observations used in this work. All
of the datasets were obtained through the large
30\arcsec$\times$30\arcsec~(LWRS) aperture. The two-dimensional {\it FUSE}
spectral images were processed using the standard CALFUSE pipeline
(version 1.8.7)\footnote{See http://www.fuse.pha.jhu.edu/analysis/pipeline\_reference.html}.
The reduction and calibration procedure includes
thermal drift correction, geometric distortion correction,
heliocentric velocity correction, dead time correction, and wavelength
and flux calibration.
The observations of each star were broken into
several sub-exposures, which were aligned by cross-correlating the
individual spectra over a short wavelength range that contained
prominent spectral features. The aligned sub-exposures were co-added,
weighting each by the exposure time. No co-addition of data from
different channels was performed; instead each channel was used
independently in the fitting procedure as a different measurement.

\section{Analysis}

In this section we discuss the analysis of atomic species O~{\small I}, H~{\small I},
Kr~{\small I}, and molecular hydrogen (H$_{2}$). Whenever possible we compare
our results with those found in the literature. The analysis of H~{\small I}
(from STIS data) and H$_2$ (from {\it FUSE} data) are described respectively in sections
\S3.1 and \S3.2. \S 3.3 discusses the analysis of O~{\small I} with other
atomic species (using STIS data).

\subsection{H~{\small I} Analysis}

One of the most important factors limiting the accurate measurement of the
{\rm O/H} ratio is the uncertainty associated with the available H~{\small I}
measurements, which are derived primarily from analyses of the damping
wings of H~{\small I} \lya. For OB stars, the uncertainties in the H~{\small I}
measurements are largely systematic, particularly due to the
complexity of the stellar continuum against which the interstellar
\lya\ is viewed. The systematic uncertainties can be
minimized, however, in some cases by using high signal-to-noise
datasets to probe further into the core of the \lya\ profile (e.g.,
Howk et al. 1999a).

We have endeavored to measure the H~{\small I} column densities towards the
stars in our sample, making use of the STIS E140H spectra discussed
above. Because the \lya\ transition covers several \AA, we
have co-added several echelle orders for our analysis. The CALSTIS
flux calibration of the E140H data shows a systematic error in that
the short wavelength portion of an order $m$ shows a larger flux than
the long wavelength portion of the order $m+1$. We have corrected
this by applying a polynomial flux correction to each order. The
correction was assumed to be the same for each order within an
observation and derived from a comparison of STIS and GHRS
data of HD218915. Unfortunately no other GHRS target was available for this comparison
but we are confident in our correction for it produces a symmetrical
profile of the \lya\ transition in each spectrum. We used a weighted average scheme 
to combine adjacent flux-corrected orders.

Given the confusing presence of stellar photospheric and wind lines in
the region near interstellar \lya, we make use of the continuum
reconstruction method (Bohlin 1975; Diplas and Savage 1994; Howk et
al. 1999a) for deriving N(H~{\small I}). This method multiplies the observed
\lya\ profile by $\exp^{(+\tau_\lambda)}$, where $\tau_\lambda$ is the 
optical depth as a function of wavelength. The H~{\small I} column density is
determined by the value of $\tau \, [\propto N(\mbox{H~{\small I}})]$ which
best reconstructs the damping wings of \lya\ to the estimated position
of the stellar continuum. The method relies on human judgment to
set the stellar continuum, which is not rigorously defined given the
undulating stellar absorption features. For each of the stars, two
authors have independently determined the $\pm$~2~$\sigma$ limits
following Howk et al. (1999a) with the final value being negotiated
between them. The $\pm$~2~$\sigma$ limits are those column densities
which showed reconstructed continua that were highly implausible.
The best fit values were then calculated to be the midpoint, assuming
symmetric errors. Finally, we find that our column
densities are, in all cases, within $1~\sigma$ of those given by Diplas \& Savage
(1994). However, the higher signal to noise
ratio of our {\it HST} data allows us to use regions closer to the line
core, leading to smaller systematic uncertainties. An example of this
procedure is shown in Figure 1, where the reconstructed continua for HD210839
are displayed.

One potential difficulty in the H~{\small I} analysis is the contamination of
the ISM \lya\ profile by stellar H~{\small I}. In particular, for stars earlier
than B2 the contamination is important for ``low'' interstellar column densities
($\log N(\mbox{H~{\small I}}) \la 20.5$; see Diplas and Savage 1994). Since
our sight lines probe long pathlengths, most of which are in the
galactic disk, we are always in the regime where $\log N(\mbox{H~{\small I}})
>21$. Therefore, the corrections for stellar \lya\ are always negligible.

We have tested the consistency of our \lya\ H~{\small I} analysis for nine of
our targets using the higher-order Lyman transitions in the {\it FUSE}
spectra of these stars. For the other stars, the blending of H$_2$
and H~{\small I} is too severe to adequately check the N(H~{\small I}) determination.
For each of these nine targets, we constructed a complete model of the line
of sight (velocity components, atomic and molecular content) which was then
compared with the {\it FUSE} profiles of the weaker H~{\small I} transitions. These
models provided two checks of our \lya\ analysis: first, they
confirmed, using the weak damping wings of Lyman-$\beta$, the total H~{\small I}
column derived from the STIS data (although with larger uncertainties);
second, modeling the high-order lines ruled out the existence of
high velocity, low-$N(\mbox{H~{\small I}})$ components which might
adversely affect our simple assumptions about the distribution of
\lya\ optical depth with wavelength (see Appendix A of Howk et al. 
1999a). This eliminated a potentially
significant bias in our H~{\small I} column density determinations caused by
hot, low column-density clouds similar to those discussed in this context by
Lemoine et al. (2002) and Vidal-Madjar \& Ferlet (2002) for sight lines through the local ISM.

\subsection{H$_2$ Analysis}

The {\it FUSE} bandpass contains some 20 ro-vibrational bands of molecular
hydrogen (see, e.g., Shull et al. 2000; Tumlinson et al. 2002) which
can be used for the determination of \htwo\ column densities.
Most of the sight lines studied here lie predominantly within the galactic disk
and have quite high molecular hydrogen column densities. 
We adopt wavelengths, oscillator strengths, and damping
constants for the molecular hydrogen transitions from Abgrall et al. (1993a)
for the Lyman system and Abgrall et al. (1993b) for the Werner system. 

Our analysis of \htwo\ with the {\it FUSE} data makes use of the $Owens$ profile fitting software
(Lemoine et al. 2002, H\'ebrard et al. 2002a) to fit the \htwo\ line profiles assuming a
single Maxwellian component. We assume that the {\it FUSE} LSF is
well described by a single Gaussian with a FWHM of 20 km~s$^{-1}$.
While this is an oversimplification of the {\it FUSE} LSF (see Wood et
al. 2002, H\'ebrard et al. 2002a), it is sufficient for our purposes.

In the course of determining the \htwo\ content, we proceed in two steps.
First, we analyse the population of the rotational levels J=2 and 3 that
contain information on the excitation temperature within the gas.
Although most of the corresponding absorption lines are deeply saturated, a few lines
are weak enough to be near the linear part of the curve
of growth, thereby allowing us to get reliable column density estimates. The profile fitting is
done simultaneously over the different absorption lines ($\approx$ 10 spectral windows).

Then, we investigate the J=0 and 1~rotational levels that represent more
than 90\% of the total \htwo\ column density. For all of the sight lines presented in
this work, the strongest J=0,1 lines show damping wings implying
$N({\rm H}_2) \ga 10^{18}$ cm$^{-2}$ (Tumlinson et al. 2002). 
The specifics of the unresolved component structure were unimportant for
these rotational levels. The main difficulty in analyzing these
lines is determining the stellar continuum. Catanzaro et al. (2001)
showed that, for early type stars with low rotational velocities, stellar
continua are relatively smooth and featureless over the v1-0 and v2-0
ro-vibrational bands. We follow Catanzaro et al. (2001) in
analyzing the J=0 and J=1 levels using the v1-0 and v2-0 Lyman bands. The J=0
transitions in these bands are centered at $\lambda \sim 1092$ and
1077 \AA. The profile fitting is simultaneously performed over 5 spectral windows corresponding
to the different channels. An example of the fit quality is shown in Figure 2.

We find that the rotational levels J=2 and 3 represent only a few percent of the
total \htwo\ column density in all lines of sight studied here. As a result, they make relatively minor
contributions to the error budget. Finally the errors are dominated by
uncertainties in the continuum placement near the strong J=0,1 \htwo\
transitions. We have estimated the contribution of the
continuum-placement uncertainties by fitting
the data with three different continua: the best continuum, the best
continuum plus 10\%, and the best continuum minus 10\%. We then
assume that the \htwo\ column densities derived with the $\pm$~10\%
continua represent $1~\sigma$ deviations from the best-fit columns.

Table 4 presents the derived \htwo\ column densities for each line of
sight, as well as the fractional abundance of molecular hydrogen,
$f({\rm H}_2)\equiv 2N({\rm H}_2)/[N({\rm H\ I})+2N({\rm H}_2)]$ and the
T$_{01}$ rotational temperature. We
have compared our H$_{2}$ results with previous analyses of the {\it FUSE}
data along two high extinction cloud sight lines studied by Rachford et al. (2002). In Table 5
we compare our results with theirs.

\subsection{O~{\small I} Analysis}

To determine the column density of oxygen along our 19 sight lines, we
use the weak O~{\small I} intersystem transition at 1355.598 \AA; the extreme
weakness of this transition allows us to avoid most of the saturation
effects seen in the other O~{\small I} lines observable by STIS and {\it FUSE}. In agreement
with Meyer (2001),we adopt
the f-value for the 1355~\AA~transition suggested by Welty et al
(1999): f=$1.16\times10^{-6}$. This value was derived from the mean
of the lifetime measurements of Johnson (1972), Wells and Zipf (1974),
Nowak, Borst and Fricke (1978) and Mason (1990) and the theoretical
branching fraction of Bi\'emont and Zeippen (1992). This value is
slightly smaller than the one previously adopted ($1.25\times10^{-6}$,
Morton 1991). When comparing our results with previous works that
make use of the older f-value, we linearly correct their column densities
to reflect our choice of oscillator strength. Our derived errors
do not include uncertainties in the f-value (which are in large part
unknown).

Although the O~{\small I} 1355~\AA~line is typically quite weak, the presence of cold components with
significant amounts of matter could cause difficulties when deriving
the O~{\small I} column densities if the lines are sufficiently narrow to be
unresolved by STIS. We used 2 different approaches to derive the O~{\small I}
column densities: profile fitting (PF) and the apparent optical depth
(AOD) method. The AOD method (Savage \& Sembach 1991, Lehner et al. 2002) yields
apparent column densities, N$_{a}$, that are equivalent to the true
column density if no unresolved saturated structure is present. If
such structure is present, then N$_{a}~<$~N, and the apparent column
density is a lower limit to the true column density. Our AOD
analysis, including estimation of the errors, follows the
procedures outlined in Sembach \& Savage (1992).

Profile fitting, in which a detailed component model of the
interstellar absorption is compared with the data, was performed with the profile
fitting code $Owens$ developed by Martin Lemoine and the {\it FUSE} french team. We assume that the STIS LSF is
well represented by a Gaussian with a FWHM of 2.67 km~s$^{-1}$ for the
spectra obtained through the $0\farcs2\times0\farcs09$ and $0\farcs2\times0\farcs2$ slits.
For the $0\farcs1\times0\farcs03$ slit
the adopted LSF is a Gaussian with a FWHM of 1.5 km~s$^{-1}$. The
errors were calculated with the $\Delta\chi^2$ technique for each of the fitted components
(H\'ebrard et al. 2002a, Lemoine et al. 2002). The profile fitting
analysis of the O~{\small I} column density was done simultaneously with other
atomic species with absorption lines visible in the STIS data. Each fit
made use of the following species to constrain the line of sight
velocity structure: Cl~{\small I} $\lambda$1347, C~{\small I} $\lambda$1276 and C~{\small I}
$\lambda$1270, S~{\small I} $\lambda$1295 and Kr~{\small I} $\lambda$1235. Results of the fits
for these species will be presented in a companion paper. Cl~{\small I} and S~{\small I}
are useful tracers of dense clouds, while C~{\small I} also traces cool
material along diffuse lines of sight. Because these atomic species
have different masses we were able to constrain the kinetic
temperature and the non-thermal broadening for each component along the
line of sight. An example of the full fit to all of the atomic
species is shown in Figure 3 for the sight line towards HD210839.
Figures 4 and 5 show the resulting fits for O~{\small I} along the 19
sight lines.

Table 6 shows the comparison between the
results obtained with the two methods, the equivalent width and the adopted N(O~{\small I}).
The adopted values are weighted averages of the two techniques, while
the adopted uncertainties are the largest of the two errors. 
Previous O~{\small I} measurements can be found in the
literature for six of our targets: HD75305, HD104705, HD177989,
HD185418, HD218915 and HD303308. Our measurements are compared with
the literature values in Table 5. With one exception our results are
consistent with previous measurements, within 1~$\sigma$. For the
sight line towards HD218915, we found that there is no need for the
saturation correction applied by Howk et al (2000), so our value is
30\% smaller than previously stated.

Meyer et al. (1998) and Cartledge et al. (2001) have
discussed the use of Kr~{\small I}, a noble gas, as a proxy for neutral hydrogen, demonstrating
the relative constancy of O/Kr in local gas. For four stars we were
able to derive Kr~{\small I} column densities, which, when compared with the
O~{\small I} column densities, offer another opportunity to compare our results with
others. Table 7 summarizes our O/Kr measurements, showing that our
measurements are consistent with previous results.

\section{ The oxygen abundance }

\subsection{ The sample }

The dataset is composed of 19 sight lines toward distant stars ranging from .8 to 5.0 kpc. The average
distance from the Sun is 2.6 kpc while the average galactocentric radius is 7.7 kpc.
We are thus probing material well beyond the local ISM. The average distance
from the Sun in our sample is about 5 times the one found in Meyer et al (1998): $\langle$d$\rangle$ = 530 pc.
The molecular fraction ranges from 3\% to 47\% while the E$_{B-V}$ ranges from
0.16 to 0.64. Although most of our sight lines are ``translucent sight lines'' as defined
in Rachford et al (2002), a handful are purely diffuse sight lines with the gas largely in the local ISM. 

As shown in Table 1, the stars studied in this work are mostly distributed in
the disk of the Galaxy with a few of them being in the low halo ($|z|$ $>$ 250 pc):
HD88115, HD99857, HD157857, HD218915, and HD177989. HD177989 is the most distant star in the
sample (5 kpc) and is, at the same time, the highest above the galactic plane with $|z|$=1032 pc.
This line of sight goes through the ejecta of the Scutum supershell (GS 018-06+44), in the Scutum
spiral arm. Previous observations carried out by Savage et al. (2001) reveal strong and broad
absorption in the lines of Si IV and C IV centered on LSR velocities of +18 and +42 km~s$^{-1}$
and weaker absorption from these ions near $-$50 and $-$13 km~s$^{-1}$. However, the neutral gas
does not show any signature of these weak high velocity components.

Among the densest sight lines in our dataset are HD210839 and HD192639 with
E$_{B-V}$=0.62 and 0.64, respectively. HD210839 is much closer to the Sun than HD192639 and has been long recognized
to have a peculiar extinction curve (Diplas \& Savage 1994). The average density is about 80
hydrogen atoms per cubic centimeter as derived from O {\small I} excited lines (Jenkins \& Tripp 2001).
New diffuse interstellar band features (DIBs) have also been reported recently toward this star~(Weselak 2001).
With the highest molecular fraction of our sample (47\%) this sight line is likely to probe dense material in
which the oxygen might be more depleted onto grains. HD192639 is part of the Cyg OB1 association located at 1.8
kpc from the Sun.
However, for this latter sight line a deep investigation of the C~{\small I} transitions with the STIS E140H echelle,
by Sonnentrucker et al. (2002) shows that there is a dense, albeit relatively small, knot along this sight line,
which has diffuse global properties (with an expected average of 11 atoms cm$^{-3}$).

Two peculiar sight lines in our sample show intermediate velocity components:
HD93205 and HD93222. Both of these stars are part of the Carina nebula and are only 400 parsec
away from each other. In a recent survey of the Carina nebula (Garcia \& Walborn 2000),
it has been demonstrated that the high and intermediate velocity structures are produced by an interaction
between the stellar winds of the earlier type stars and immediately surrounding interstellar material.
Small angular variations in Na~{\small I} and Ca~{\small II} abundances are typical in this nebula
and we shall expect the same for the oxygen abundance if the absorption is mainly due to the
surrounding clouds.

\subsection{Results}

Table 6 summarizes all the measurements performed for
\oh. Using the 19 stars of the sample, we find a mean abundance of
\oh~= 408 ppm with an error in the mean of 13 ppm. The error in
the mean we report here is the weighted average variance which is computed with
the individual error estimates and the scatter around the mean. We also
computed the {\it a priori} estimate of the error in the mean using only the individual
error estimates. While this usually should give a smaller value, we find 14 ppm
and this result strongly suggests that some of our errors have been
overestimated. The weighted average variance is certainly
a more reliable estimate for the error in the mean in this case. Throughout this section, we
also use the weighted average variance as the best estimate of the error in the mean
when comparing different samples.
The straight standard deviation is 59 ppm and represents only a 15\% deviation from the mean.
The reduced $\chi^2$ of 0.9 indicates that the scatter is consistent
with the errors assuming a Gaussian parent distribution around a constant value.

If we only consider the low halo targets ($|z| > 250$ pc), we have 5 stars with 
a mean of \oh~= $394 \pm~24$ ppm. These sight lines to halo stars appear to be 
statistically consistent with the rest of the sample; this is due to 
the fact that these sight lines are probing mostly disk material. In particular,
in the case of HD177989, we have seen that none of the typical velocity structures
associated with high-z features along this sight line are detected in the neutral gas.

The fact that we find a homogeneous oxygen abundance throughout the disk contrasts
with the large scatter reported in stellar atmospheres studies (Smartt \& Rolleston 1997,
Primas et al. 2001). It could be that this scatter vanishes as we integrate the column
density over long pathlengths. On the other hand, our scatter is consistent with the
rms scatter found in the H~{\small II} survey by Deharveng et al. (2000) ($\approx$ 22 \%)
and is also consistent with the picture of an efficient mixing of oxygen in the galactic disk as shown by
theoretical calculations (see Roy \& Kunth 1995). 

The constancy of the oxygen abundance holds for a large range of molecular fractions
as shown in Figure 6. It is remarkable to observe that, although many of the
targets in our sample have molecular fractions similar to the ones reported in Cartledge et al. (2001),
we do not report any trend toward an enhanced oxygen depletion.
One possible explanation is that the sight lines in our sample
are made of a mixture of diffuse and dense clouds and are indeed dominated by the diffuse
components in most cases, contrary to the most reddened sight lines in their study
which are spatially associated with dark cloud complexes. 
However, in the case of the sight line toward HD210839,
which presents both high E$_{B-V}$ and large density (80 hydrogen atoms per cubic centimer),
we still do not detect the hint of an enhanced oxygen depletion reported by Cartledge et al. (2001).

In order to test how much the oxygen abundance is varying with distance and density
we combine our measurements with measurements from the recent literature. The 13 sight lines
from Meyer et al. (1998) are similar to the sight lines in our sample (referred as sample A) in
the sense that they are mostly diffuse and are not expected to show O~{\small I} enhanced depletion
onto grains. Noteworthy, the average distance in sample A is about one order of magnitude
larger than in the Meyer et al. (1998) sample.
The comparison of the mean \oh~also reveals a systematic
16\% underabundance of oxygen within 500 pc, as illustrated in Figure 7.

By combining Meyer et al. (1998) 
values with ours, we get a sample of 32 sight lines (sample B) in which the distances range from 150 pc
to 5 kpc. We find a mean \oh~of 377 ppm with an error in the mean of 10 ppm.
Finally, we include 5 measurements performed through denser media by Cartledge et al. (2001):
HD27778, HD 37021, HD37061, HD147888 and HD207198. While the distances of these 5 stars are comparable to
the distances found in sample A, the sight lines are different in nature being probably a lot denser.
This final sample (sample C) is made of 37 sight lines with a wide variety of properties: the total hydrogen
column density ranges from 1.5$\times$10$^{20}$ cm$^{-2}$ to 62$\times$10$^{20}$ cm$^{-2}$,
log f(H$_2$) from -5.21 to -0.2, the reddening from 0.09 to 0.64 and the distance to
the Sun from 150 pc to 5 kpc. Table 8 summarizes the oxygen abundance found in the different samples.
In particular, we see that in sample C, although the scatter is the largest, the standard deviation
is only of the order of 20\% around the mean of 362 ppm.















From the 19 measurements of the total hydrogen in sample A, we derive an
average  $\langle$ H$_{tot}$/E$_{B-V}$ $\rangle$ of 6.3$\times$10$^{21}$ cm$^{-2}$ mag$^{-1}$ with a standard
deviation of 9$\times$10$^{20}$ cm$^{-2}$ mag$^{-1}$ (Figure 8). This value is fully
consistent with the known galactic averages $\langle$ H$_2$/E$_{B-V}$ $\rangle$ = 5$\times$10$^{20}$ cm$^{-2}$ mag$^{-1}$ (Dufour
et al. 1982), $\langle$ H~{\small I}/E$_{B-V}$ $\rangle$ = 4.93$\times$10$^{21}$ cm$^{-2}$ mag$^{-1}$ (Diplas \& Savage 1994)
implying $\langle$ H$_{tot}$/E$_{B-V}$ $\rangle$ = $\langle$ H~{\small I}/E$_{B-V}$ $\rangle$ + 2 $\langle$ \htwo/E$_{B-V}$ $\rangle$
= 5.93$\times$10$^{21}$ cm$^{-2}$ mag$^{-1}$.
The standard deviation we derive for this ratio is relatively small and is
an indication that the total hydrogen content and the dust are well correlated
in our sample. Finally, we point out that oxygen column density and reddening seem to be
comparable tracers of the total hydrogen column density over a wide range of distances and
densities.

\section{Discussion}

\subsection{Dust grains}

The role of oxygen is fundamental in modern dust models. It is known to be
abundant in the cores of silicate dust grains (Hong \& Greenberg 1980).
In particular, the maximum amount of oxygen atoms
tied up into grains is limited by the abundance of metals with which oxygen bonds (O:Si:Mg:Fe)
$\approx$ (24:1:1:1). A theoretical upper limit to the number of
oxygen atoms that can be incorporated into grains is then calculated assuming
that all grains are made of silicates (Cardelli et al. 1996): (O/H)$_{\rm{dust}}$ $\le180$ ppm.
It was long believed that this value was too small to reconciliate
the gas phase ISM oxygen abundance with the solar value. However, recent measurements
of the solar oxygen abundance by Holweger (2001) and Allende et al. (2001) have revised
the solar abundance downward significantly. Allende et al. (2001) derived a mean solar oxygen
abundance of 490~$\pm$~59 ppm while Holweger (2001) derived a similar value
(although with larger errors) with $({\rm O/H})_\odot$= 545 ppm $\pm$ 100 ppm. In the following we
adopt a straight average of their values as the solar abundance : $({\rm O/H})_\odot = 517 \pm~58$ ppm.
This value is now in good agreement with the maximum total abundance (gas+dust)
derived using our gas-phase {\rm O/H} measurements combined: 
\oh~+ (O/H)$_{\rm{dust}}~\la 408 \ {\rm ppm} \, + \, 180 \ {\rm ppm} \, \la 588 \, {\rm ppm}$.
Thus it would seem that the total interstellar oxygen abundance is now
consistent with the best solar oxygen abundance.

If we assume that the solar value reflects the total oxygen abundance in the
ISM and that the total oxygen abundance is constant, we can then estimate the average dust oxygen content for our sight lines.
We find $\langle$(O/H)$_{\rm{dust}}\rangle$ = 109 ppm with a standard deviation of
about 60 ppm. Since we know the maximum amount of oxygen atoms that can be tied up into silicate grains
(O/H)$_{\rm{dust}}$ = 180 ppm), then, from our survey, we find that $\approx$ 60\% of the grains are
resilient against destruction by shocks on average, with one extreme sight line showing nearly 100\% 
destruction and others showing no destruction at all. Along the sight line toward HD93222, which probes material
surrounding
the star (see \S4.1), most of the oxygen atoms have been released in the gas phase. This can be
interpreted as the effect of the interaction between the stellar wind and the nearby ISM. On the
other hand, the three most oxygen depleted sight lines in sample C are from the Cartledge et al. (2001) survey
in the dense ISM, consistent with an enhanced depletion into dust grains.
Finally, this is also consistent with the theoretical expectation that only 10 to 15\% of the grains are
destroyed at each passage of a 100 km~s$^{-1}$ shock in the warm ISM
(Tielens et al. 1994, McKee et al. 1987). Such shocks are believed to be responsible for most
of the grain destruction and could explain the observed population of resilient dust grains.
These shocks could also explain the variations in oxygen bearing dust.
Several other indirect pieces of observational evidence support the view
that dust grains are long-lived, among them
the abundance studies toward the diffuse halo clouds by Sembach \& Savage (1996) and the
detection of dusty structures in the halo of spiral galaxies by Howk \& Savage (1999b). Both
observations imply that a fraction of the grains can survive for a long time before being destroyed
by thermal sputtering in the halo.

\subsection{ Local oxygen deficit }

Meyer et al (1998) showed that the local oxygen abundance is remarkably constant. 
Our data, probing a wider range of environments at larger distances, show much the same.
However, there is a systematic difference of 16\% between the two samples that needs to be discussed in
the context of a possible enhanced depletion within a few hundred parsecs from the Sun.

Previous studies of the local ISM have found some evidences for a global elemental deficit.
Krypton abundance has been measured toward 10 stars in the local
ISM by Cardelli et al. (1997) and a depletion of 1/3 was found when compared to the solar abundance.
Because it is a noble gas, krypton is not expected to be depleted into the dust phase and this
observation alone is a strong argument against the Sun being an appropriate cosmic
standard for the nearby ISM. The same has been observed with sulfur, which is not supposed
to be depleted in dust grains. The sulfur abundance derived from nearby B stars is about 2/3 of the observed ISM
abundances beyond 2 kpc (Fitzpatrick \& Spitzer 1993, 1995, 1997 and Harris \& Mas 1986) and 2/3 of the solar sulfur abundance.
The same depletion pattern is seen for oxygen when the solar value is compared to the nearby B stars
while the comparison with distant B stars does not shed much light on the subject because of the large scatter
(Smartt \& Rolleston 1997). More generally, abundances have been measured in nearby B stars and
young F and G field stars (see the summary by Snow \& Witt 1996) and are consistently below solar
for all measured elements. These star samples are representative of the current local ISM and clearly
point toward a local elemental deficit.

Meyer et al. (1998) have reviewed the possible scenarios to explain such
a depletion pattern: presolar nebula enrichment, radial migration of the Sun and infall of metal-poor gas.
The oxygen abundance we derive from outside the local ISM can be used as a test for the latter scenario.
As far as the oxygen abundance is concerned, the presolar nebula enrichment and the Sun migration might
be
discarded because of the apparent consistency between the solar value and the total oxygen abundance of the distant
ISM. The recent infall of metal-poor material seems to be consistent with our
present observation that the oxygen gas-phase abundance is slightly higher beyond 1 kpc.
This is expected to be the case if the Gould belt has been created by an infalling cloud (Comeron \& Torra 1994).
Infall is considered as an important component of galactic chemical evolution
and is supported by evidence for the infall of low-metalicity clouds onto the Milky Way (Richter et al. 2001).
In this scenario the dilution should have affected all the elements the same way and
the local oxygen enhanced depletion is expected to be the same as observed in krypton and sulfur: about one third.

However, the difference between our value, \oh~= $408$~$\pm$~13 ppm beyond 1 kpc
and the value found by Meyer et al. (1998), \oh~= $343$~$\pm$~15 ppm within 1 kpc, is only one sixth. A possible
explanation for a reduced dilution of oxygen in the local ISM is the destruction of a
fraction of the dust grains that resulted from the impact of a low metalicity cloud, at an estimated
velocity of 100 km~s$^{-1}$ (Comeron \& Torra 1994).
However, in this scenario, we should expect that many metals were released in the gas-phase as well.
It is beyond the scope of this paper to explore the different infall scenarios that might be consistent with
the observations.

\subsection{The galactic gradient}

Some of the stars in our sample are sufficiently distant that the measurements might be affected by large scale
abundance gradients in the ISM. Of course the gradient is in part masked by the
integrated nature of our absorption lines and by the intrinsic scatter between sight lines.
An extreme example of intrinsic scatter between the data points is given by the couple HD93222/HD93205.
Both are members of the Carina nebula (Garcia \& Walborn 2000) and show intermediate velocity
structures in their spectra due to stellar winds interacting with surrounding material.
It is well known that in this nebula the spatial variation of Na~{\small I}
as well as Ca~{\small II} can be as large as 50\%. Although they are within 400 pc of each other, the oxygen abundances
we derive differ by as much as 43\%. However the typical scatter between our data points is only
of the order of 15\% and this must be compared to the gradient effect expected in our data.

Numerous papers have discussed the distribution of heavy elements in 
spiral galaxies (see the review by Henry \& Worthey 1999 and references therein).
The published values by Deharveng et al. (2000), using H~{\small II} regions, and by Smartt \& Rolleston (1997),
using B stars, give the range of expected values for the oxygen gradient $\alpha$:
-0.07$\pm~0.01$ dex kpc$^{-1}$ $<~\alpha~<$ -0.039$\pm~0.005$ dex kpc$^{-1}$.
It is difficult, if not
impossible, to use the velocity of the absorbers to deduce a precise location within the disk assuming the
most recent galactic rotation curve. The reason is that, except in a few cases, the expected velocities in the
local standard of rest along our sight lines are within the typical scatter (around 7 km$s^{-1}$).
Thus, we choose to show the oxygen abundance as a function of the location of the background star, expecting
the gradient to show up as well.
Figure 9 allows the comparison of the extreme $\alpha$ values with our data (excluding the peculiar sight lines
HD93222 and HD93205). It turns out that the scatter of the points is largely dominated by the
measurement errors while no large scale trend emerges at all.

\section{ Summary }

In this paper we have presented a survey of the oxygen gas-phase abundance using STIS/{\it FUSE} data
toward 19 stars with the purpose of probing the oxygen abundance far away
in the disk. This work follows from previous surveys in the local ISM and is possible
due to the {\it FUSE} detector's sensitivity, which permits measurements of the H$_2$ content
through long pathlengths. A summary of our results is as follow:

1.~We demonstrate that the oxygen abundance shows little variation in various environments and
various locations in the disk. We find a mean ratio of \oh~= 408 ppm with an error in the mean
of 13 ppm. This is 19\% higher than the mean previously derived in the local ISM by Meyer et al. (1998).
We argue that a recent local infall of metal-poor material is not excluded as an explanation for the difference.

2.~The oxygen abundance derived from our sample combined with previous surveys (37 sight lines)
is consistent with both the oxygen abundance of young G \& F stars ($\langle {\rm O/H} \rangle
= 456\pm~156$ ppm; Sofia \& Meyer 2001) or the newly revised solar value ($\langle {\rm O/H}
\rangle = 517\pm~58$ ppm; Holweger 2001, Allende et al. 2001) once dust is taken into account.

3.~Dust grains in the ISM are probably made of grain cores and we estimate that at least 60\% of the grains
have to be resilient in order to explain the difference between ISM and stellar abundances.
















It is likely that the O~{\small I} column density can be used as a surrogate tracer of the total neutral hydrogen
column density, at least in a statistical sense, in diffuse ISM abundance studies. Although above we
have emphasized the 19\% difference between the mean \oh~found in this study and the lower value
found for the LISM by Meyer et al. (1998), the most striking result from both studies is the lack of
large deviations from the mean. The standard deviation in the sample reported here is 15\%. If the Meyer
et al. (1998) and the Cartledge et al. (2001) data are included, the standard deviation increases
to only 20\% for a wide variety of diffuse ISM conditions. The use of O~{\small I} as a tracer of the total hydrogen
column density could be of use in a variety of chemical abundance studies. As an example, consider the
D/H studies, presently under way with the {\it FUSE} satellite. In many cases, determining accurate H~{\small I}
column measurements requires using {\it HST} or {\it EUVE} data, which may not be available. In other cases,
because the column densities of D~{\small I} and H~{\small I} differ by about five orders of magnitude, accurate measurements
of both species may be precluded by special conditions associated with a particular sight line. Preliminary
studies of D/O ratios using {\it FUSE} measurements of D~{\small I} and O~{\small I} are encouraging (Moos et al. 2002, and references therein, H\'ebrard et al. 2002b). Additional studies are also needed to determine the special cases in which there
are large deviations in the \oh~and identify the special conditions under which such deviations
might exist (e.g Hoopes et al. 2003).

\acknowledgments

We are grateful to Paule Sonnentrucker for sharing with us information
on the lines of sight toward HD185418 and HD192639. We also want
to thank Dr. Ken Sembach for fruitful advice. We are pleased to acknowledge
comments by Dr. Stefan Cartledge, Dr. James Lauroesch and Dr. David Meyer
leading to several improvements on this paper. We thank also the referee
for his thoughtful remarks.
This work is based on data obtained for the Guaranteed Time Team by the NASA-CNES-CSA
{\it FUSE} mission operated by the Johns Hopkins University. Financial support
to U. S. participants has been provided in part by NASA contract NAS5-32985 to
Johns Hopkins University. Support for french participation in this study has been
provided by CNES. Based on observations made with the NASA/ESA Hubble Space
Telescope, obtained from the Data Archive at the Space Telescope Science Institute,
which is operated by the Association of Universities for Research in Astronomy, Inc.,
under NASA contract NAS5-26555. These observations are associated with proposal
8241. This work has been done using the profile fitting procedure Owens.f developed
by M. Lemoine and the French {\it FUSE} team.







\appendix























\clearpage 

\begin{figure}
\epsscale{1.}\plotone{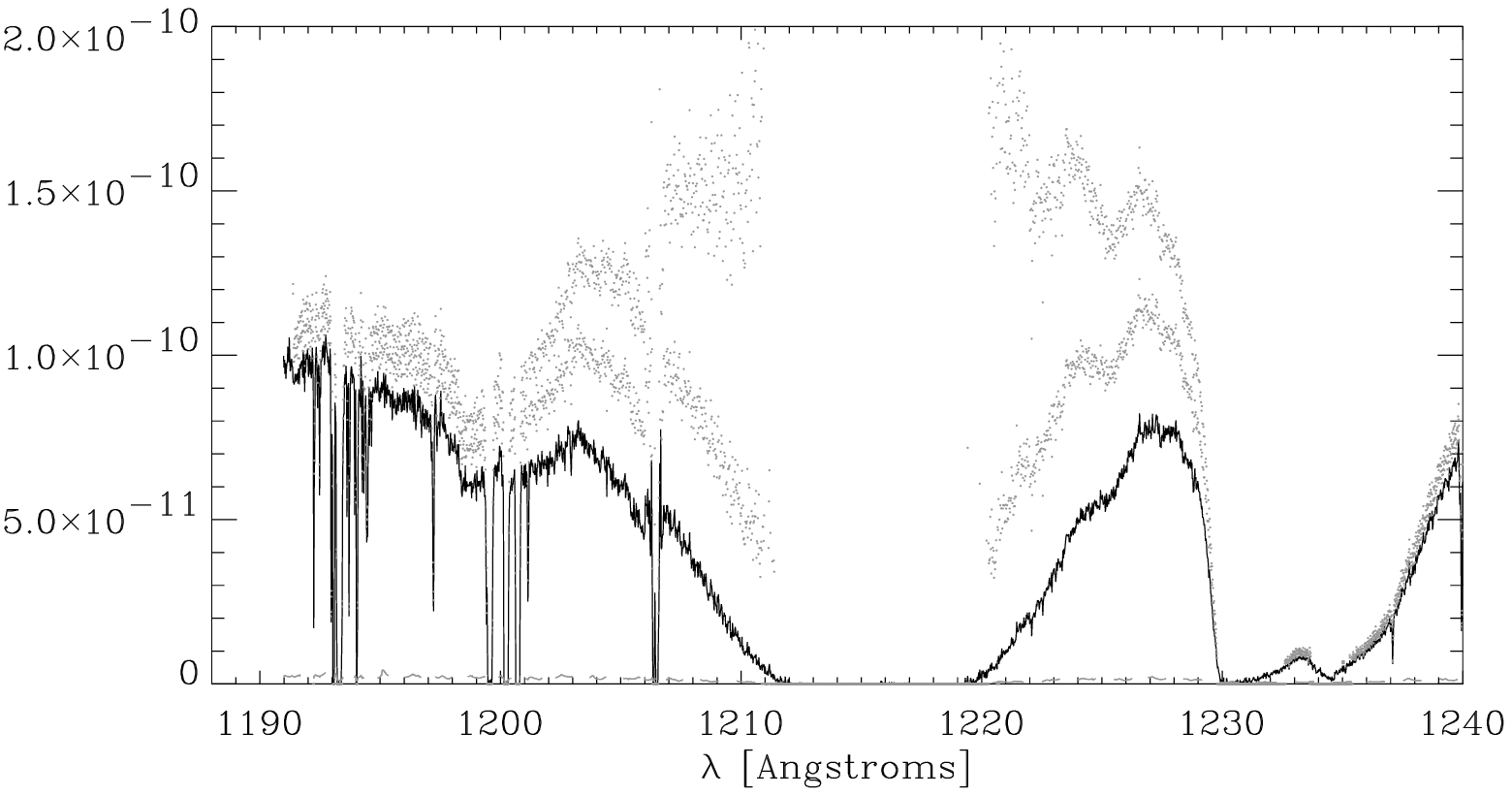}
\caption{Example of \lya\ continuum reconstruction for HD210839.
The black continuous line represents the data, the gray points are the
reconstructed continua. Each of these continua is at least
two sigma above and below from the true continuum. Note the steep absorption
feature on the red wing of \lya\ associated with stellar wind.}
\end{figure}

\clearpage 

\begin{figure}
\epsscale{1.}\plotone{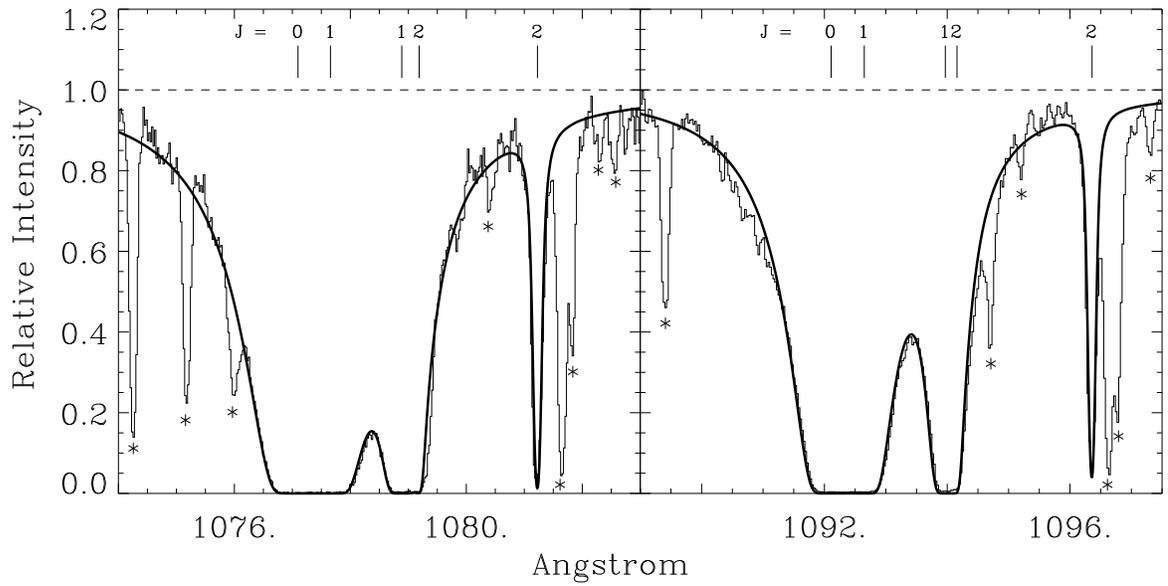}
\caption{Fits of the low rotational levels (J=0, 1, 2) in the v2-0 (right) and v1-0 (left) bands 
for HD210839. For each of the spectral windows we used all the available {\it FUSE} segments (see Moos et al.
2000): LiF1A, SiC1A
and SiC2B for v2-0; LiF2A and SiC2B for v1-0. Higher rotational transitions are indicated by
the asterisks, they have been excluded from the fit.}
\end{figure}

\clearpage 

\begin{figure}
\epsscale{0.4}\plotone{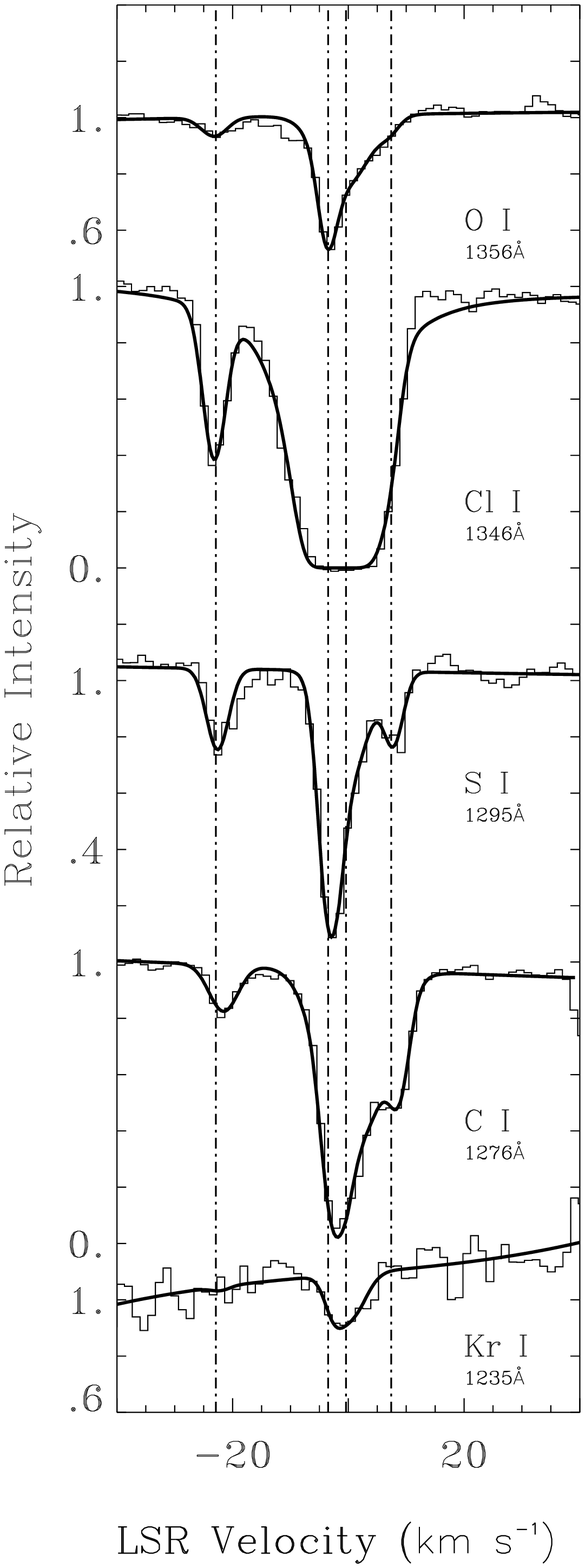}
\caption{This plot shows the fits of the atomic species for HD210839.
This line of sight is an example of complex velocity structure with
four components. Note that the signal to noise is low around the Kr~{\small I} line
due to the stellar wind absorption feature around 1235 \AA.}
\end{figure}

\clearpage 

\begin{figure}
\epsscale{0.7}\plotone{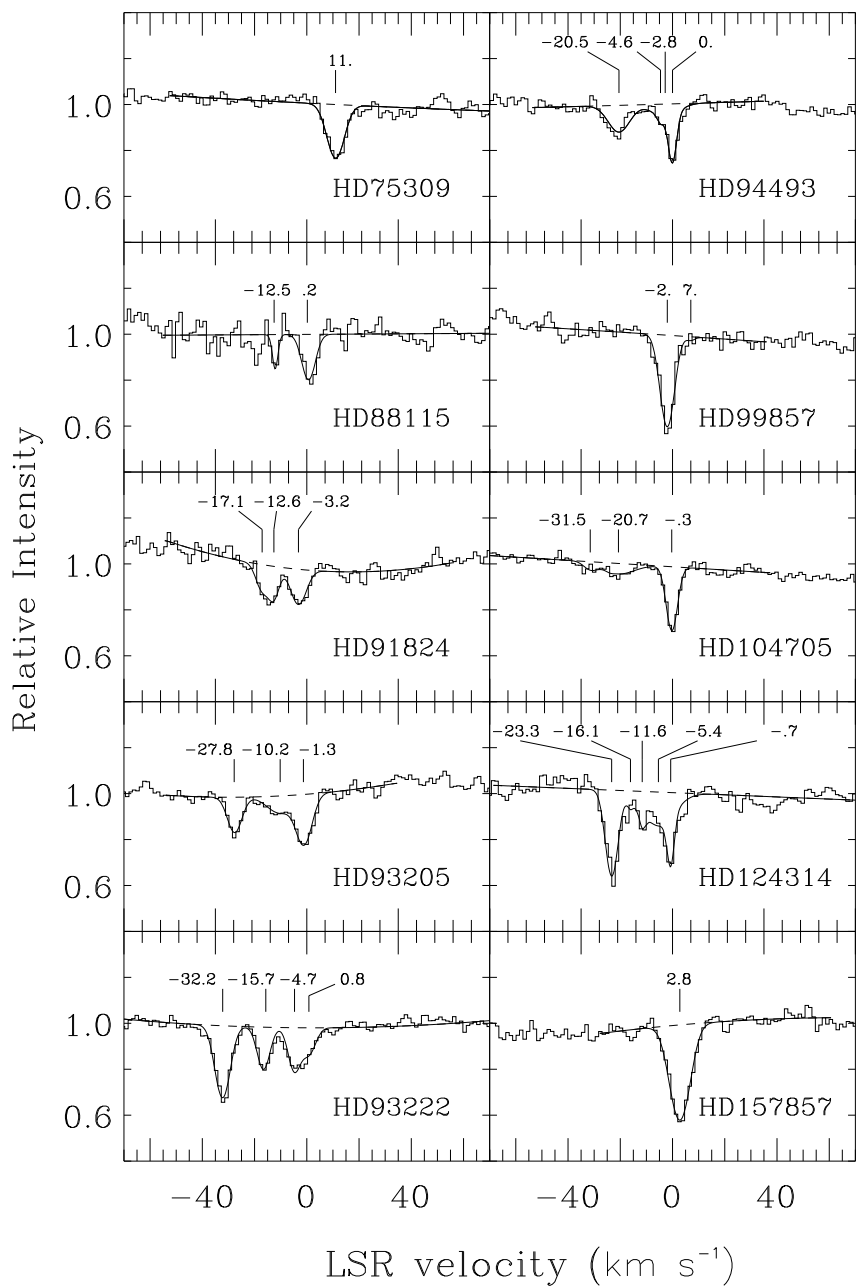}
\caption{ Profile fitting of the O~{\small I} $\lambda$1355 line. The numbers with
ticks above the plot represent the velocity of each component. For each of these fits,
we performed simultaneous profile fitting of C~{\small I} $\lambda$1276, C~{\small I} $\lambda$1279,
Kr~{\small I} $\lambda$1235, Cl~{\small I} $\lambda$1246 and S~{\small I} $\lambda$1295. Thus the information about the
velocity structure and the b-values are well constrained. In the case of HD88115, the
weak blue component is also detected by Jenkins \& Tripp (2001).}
\end{figure}

\clearpage 

\begin{figure}
\epsscale{0.7}\plotone{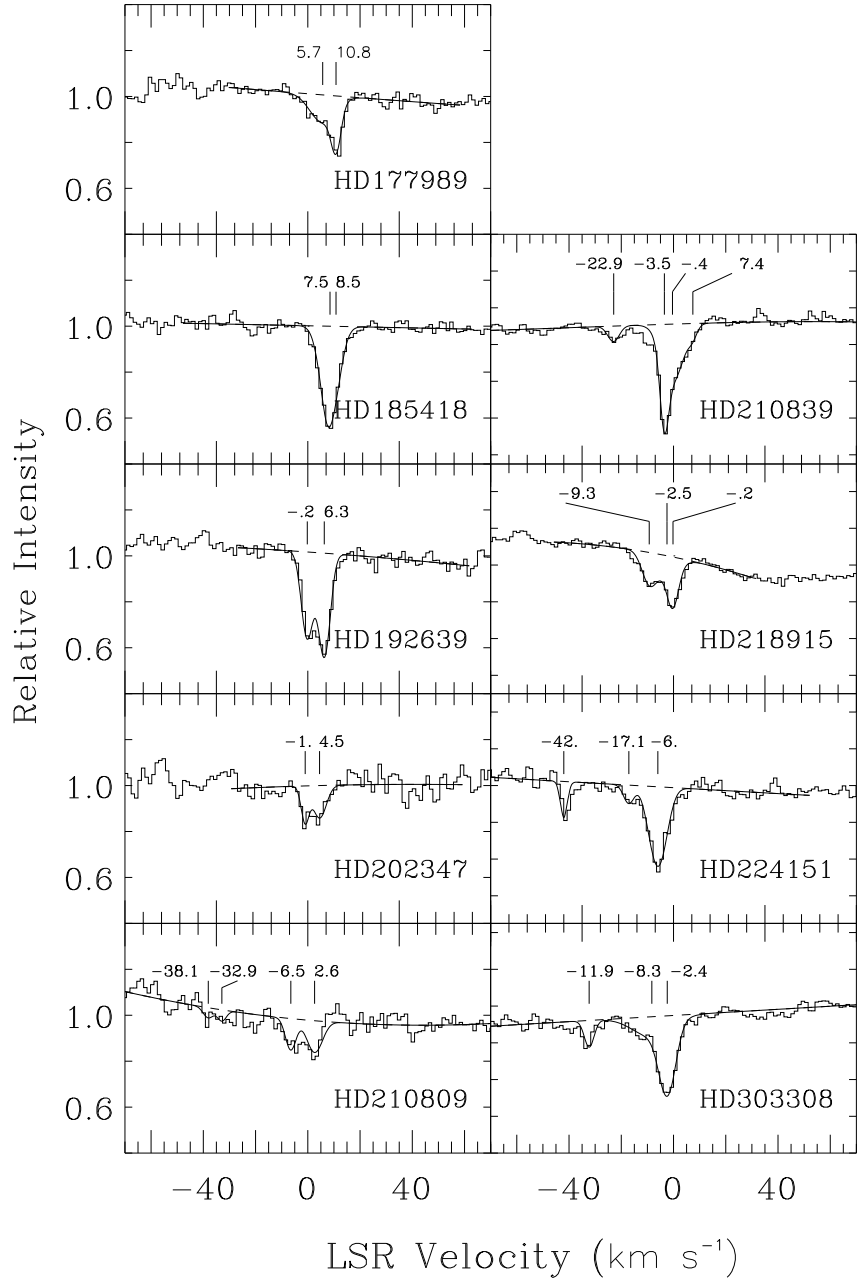}
\caption{ Same comments as Figure 4. Note the noisy spectrum
of HD210809 and the two weak O~{\small I} components (detected in Cl~{\small I}, C~{\small I} and S~{\small I})
dominated by the noise. This produces large errors for this line of
sight.}
\end{figure}

\clearpage

\begin{figure}
\epsscale{1.}\plotone{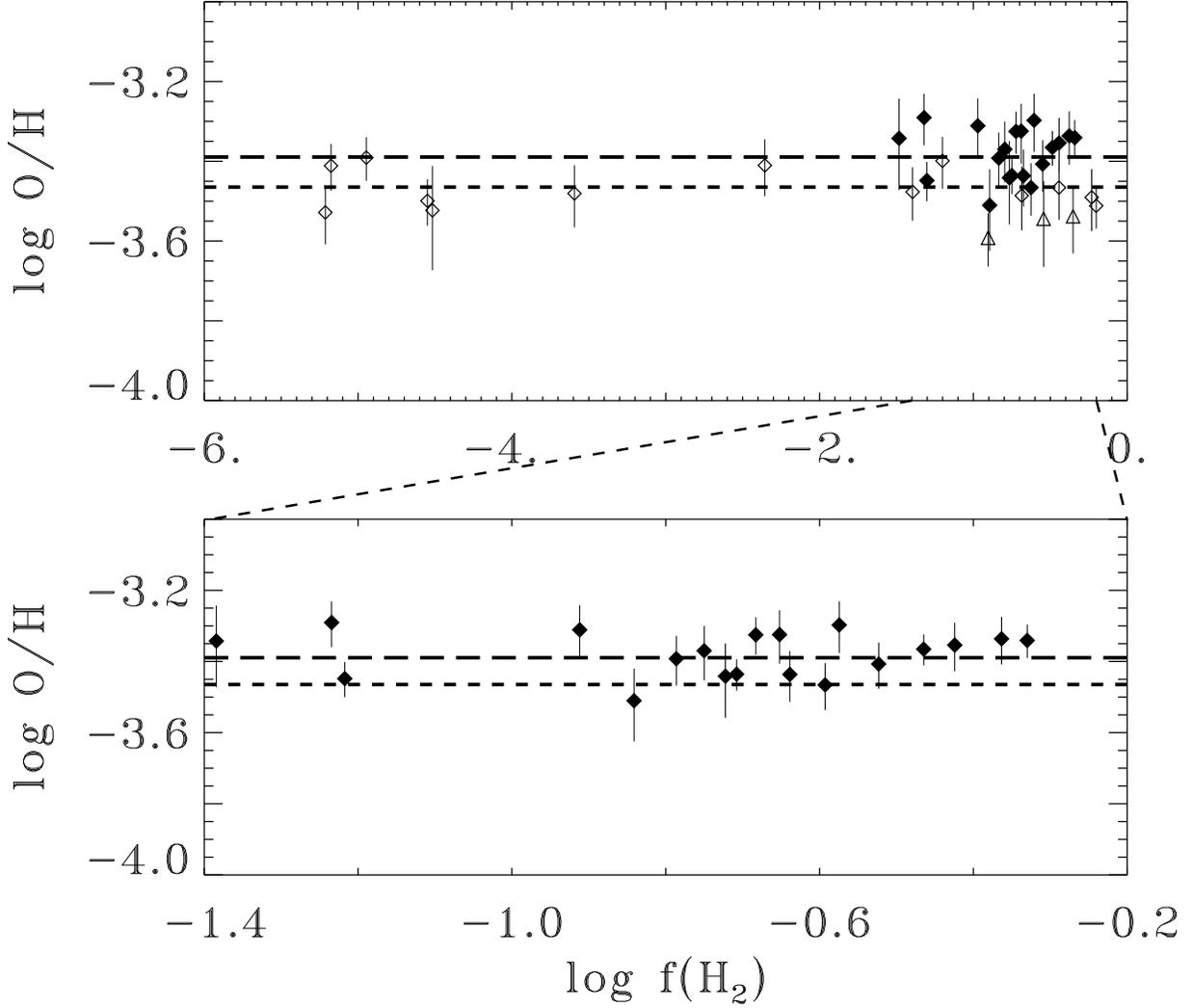}
\caption{The dashed line is the
mean of the Meyer et al. (1998) sample: $({\rm O/H})_{gas}$=343 ppm. The long dashed line
is the mean of our 19 sight lines: $({\rm O/H})_{gas}$=408 ppm. The error bars are 1~$\sigma$. Filled diamonds
are our new measurements, open diamonds are Meyer et al. (1998) and
open triangles are the Cartledge et al. (2001) measurements for which the
molecular fraction is known. No enhanced depletion
is seen for up to 47\% of hydrogen atoms in H$_2$ within our
sample. However, stars used in Cartledge et al. (2001) are both
closer to the Sun and with a higher H~{\small I} column density. They cross
denser media in average and appear to be consistently below the mean.
}
\end{figure}

\clearpage 

\begin{figure}
\epsscale{1.}\plotone{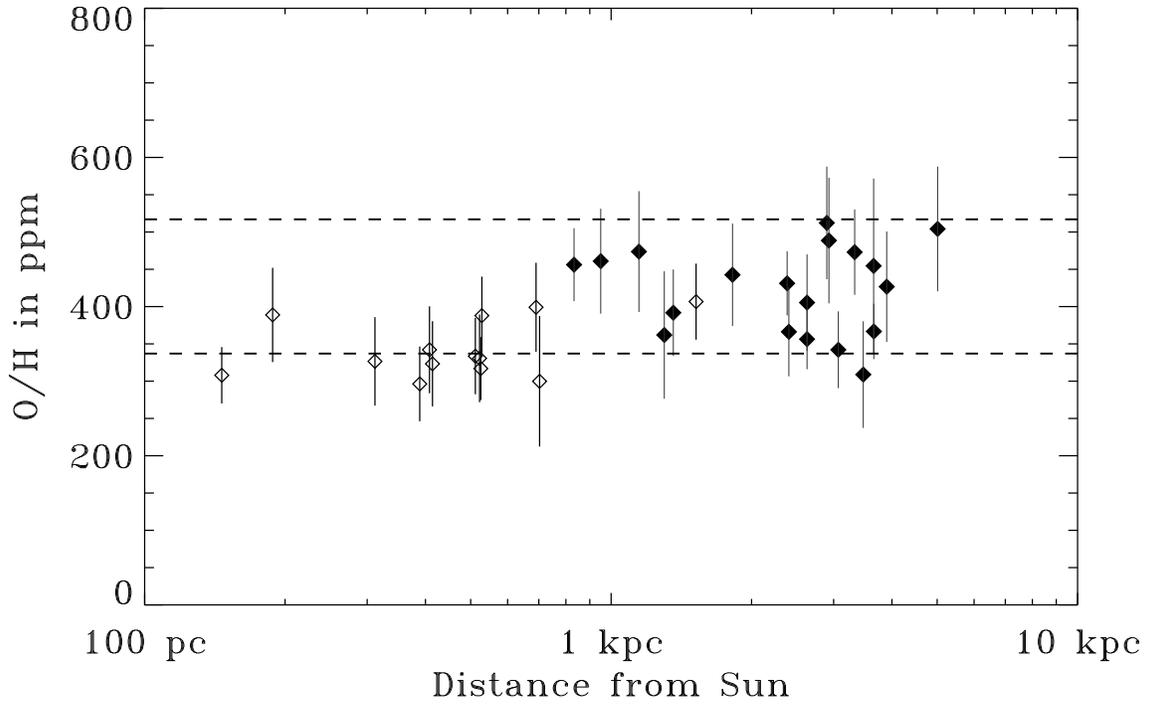}
\caption{{\rm O/H} versus distance from the Sun. Filled diamonds are
our new measurements in the distant ISM, open diamonds are local
measurements from Meyer et al. (1998). The upper dashed line
represents the total oxygen abundance (solar value) and the lower dashed line represents
what is left when the maximum amount of oxygen (180 ppm) is depleted into grains.
All the measurements are expected to lie within these two limits. However, we note that, 
for pathlengths smaller than 1 kpc, points cluster around the lower limit
and a little below.}
\end{figure}

\clearpage 

\begin{figure}
\epsscale{1.}\plotone{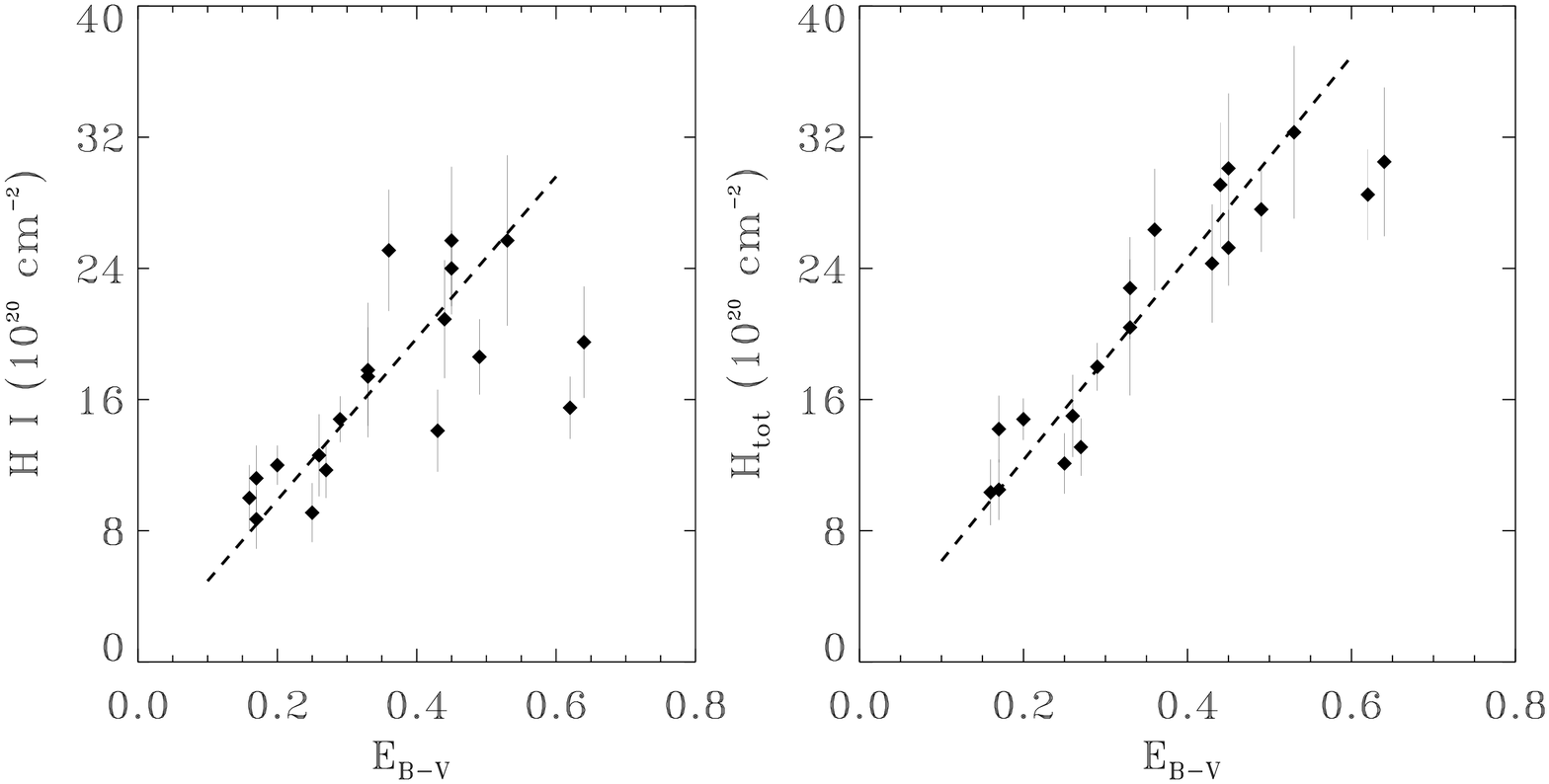}
\caption{The left panel shows the correlation between H~{\small I} and E$_{B-V}$.
We find H~{\small I}/E$_{B-V}$ = 4.9$\times$10$^{21}$ cm$^{-2}$ mag$^{-1}$ and a standard
deviation of about 25\%. The correlation
is much better between H$_{tot}$ and E$_{B-V}$ with a ratio of 6.3$\times$10$^{21}$
cm$^{-2}$ mag$^{-1}$ and a standard deviation of 15\% (right panel). Dust and 
gas are strongly correlated in our sample of diffuse sight lines.}
\end{figure}

\clearpage 

\begin{figure}
\epsscale{1.}\plotone{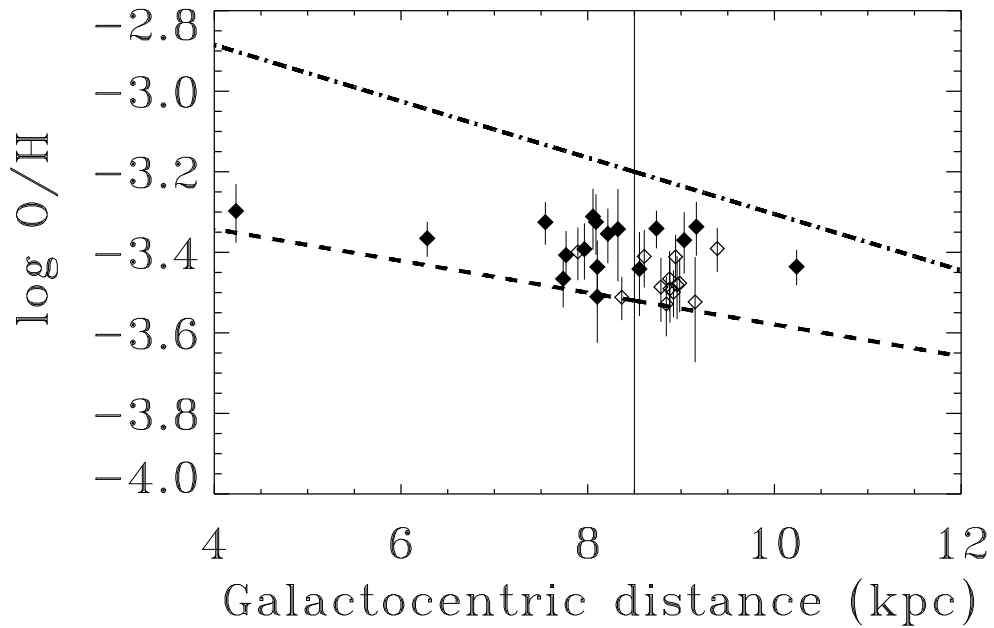}
\caption{{\rm O/H} versus galactocentric distances. The upper dashed line
represents the expected gradient from B star surveys (Rolleston
et al 1997). The lower dashed line is the expected gradient from
the H~{\small II} survey by Deharveng et al. (2000). The location of the absorbers
could be anywhere between the given stellar galactic radius and 8.5 kpc. All our
values (filled diamonds) and the values from Meyer et al. (1998) (open diamonds)
are within the range of the expected gradients. We excluded
from this plot the two sight lines toward the Carina nebula (see \S4.1). No trend
is detected as a function of the stellar location. The vertical axis
indicates the Sun's galactocentric radius.}
\end{figure}

\clearpage 

\begin{table}
\begin{footnotesize}
\begin{center}
\caption{Stellar properties.}
\begin{tabular}{ccccccc}
\tableline\tableline
Name& d & z & l & b & $E_{B-V}$&Spectral type\\
& [pc] & [pc] &&&\\
\tableline
HD75309\tablenotemark{a}~\tablenotemark{b}& 2405& 80 &265.9 & -1.9 &0.27&BI IIp  \\
HD88115 & 3654& 350 & 285.3 & -5.5 &0.16&B1 Ib/II\\
HD91824 & 2930& 5   & 285.7 & +0.1 &0.27&O6\\
HD93205 & 2630& 32  & 287.6 & -0.7 &0.45&O3 V\\
HD93222\tablenotemark{c} & 2900& 50 &287.7 & -1.0 &0.36&O7 III\\
HD94493 & 3328& 70 & 289.0 & -1.2 &0.20&B0.5 Iab/Ib\\
HD99857 & 3070& 267& 294.8 & -5.0 &0.33&B1 Ib\\
HD104705& 3900& 20 & 297.4 & -0.3 &0.26&B0 III/IV\\
HD124314& 1150& 8  & 312.7 & -0.4 &0.53&O7\\
HD157857& 2380& 548& 13.0 & +13.3 &0.49&O7\\
HD177989& 5010&1042& 17.8 & -12.0 &0.25&B2 II\\
HD185418\tablenotemark{c}& 950 & 33 & 53.6 & -2.0 &0.43&B0.5 V\\
HD192639& 1830& 48 & 74.9 & +1.5 &0.64&O7\\
HD202347\tablenotemark{c}& 1300& 45 & 88.2 & -2.0 &0.17&B1 V\\
HD210809& 3470& 188& 99.8 & -3.1 &0.33&O9 Ib\\
HD210839& 840 & 38 & 103.8 & +2.6 &0.62&O6 Iab\\
HD218915& 3660& 436& 108.1 & -6.9 &0.29&O9.5 Iab\\
HD224151& 1360& 110& 115.4 & -4.6 &0.44&B0.5 III\\
HD303308& 2630& 27 &287.6 & -0.6 &0.45&O3 V\\
\tableline
\tablenotetext{a}{Distance from Kaltcheva and Hilditch (2000).}
\tablenotetext{b}{Reddening and spectral type from Reed (2000)}
\tablenotetext{c}{Distances, reddenings and spectral types from Savage et al. (1985).}
\tablecomments{Distances, reddening and spectral types are from 
Diplas and Savage (1994) unless otherwise noted. The typical uncertainty in the distance is
30\%.}
\end{tabular}
\end{center}
\end{footnotesize}
\end{table}

\begin{deluxetable}{lcccc}
\tablecolumns{4}
\tablewidth{0pc}
\tablecaption{Summary of {\it HST} observations\tablenotemark{a}.}
\tablehead{\colhead{Name} & \colhead{Dataset} & \colhead{Date} & \colhead{Exp. time (sec)} & \colhead{Aperture} \\
}
\startdata
HD75309  & O5C05B010 & 03-28-99& 720  &  $0\farcs2\times0\farcs2$ \\
HD88115  & O54305010 & 04-19-99& 1300 &  $0\farcs1\times0\farcs03$ \\
HD91824  & O5C095010 & 03-23-99& 360  &  $0\farcs2\times0\farcs2$ \\
HD93205  & O4QX01010 & 04-20-99& 1200  &  $0\farcs2\times0\farcs09$ \\
         & O4QX01020 & 04-20-99& 780   &  $0\farcs2\times0\farcs09$ \\
HD93222  & O4QX02010 & 12-28-98& 1680  &  $0\farcs2\times0\farcs09$ \\
         & O4QX02020 & 12-28-98& 1140  &  $0\farcs2\times0\farcs09$ \\
HD94493  & O54306010 & 04-22-99& 1466  &  $0\farcs1\times0\farcs03$ \\
HD99857  & O54301010 & 02-21-99& 1307  &  $0\farcs1\times0\farcs03$ \\
HD104705 & O57R01010 & 12-24-98& 2400  &  $0\farcs2\times0\farcs09$ \\
HD124314 & O54307010 & 04-11-99& 1466  &  $0\farcs1\times0\farcs03$ \\
HD157857 & O5C04D010 & 06-03-99& 720  &  $0\farcs2\times0\farcs2$ \\
HD177989 & O57R03020 & 05-28-99& 2897 &  $0\farcs2\times0\farcs09$ \\
HD185418 & O5C01Q010 & 11-12-98& 720  &  $0\farcs2\times0\farcs2$ \\
HD192639 & O5C08T010 & 03-01-99& 1440  &  $0\farcs2\times0\farcs2$ \\
HD202347 & O5G301010 & 10-09-99& 830  &  $0\farcs1\times0\farcs03$ \\
HD210809 & O5C01V010 & 10-30-98& 720 &  $0\farcs2\times0\farcs2$ \\
HD210839 & O54304010 & 04-21-99& 1506 &  $0\farcs1\times0\farcs03$ \\
HD218915 & 057R05010 & 12-23-98& 2018 &  $0\farcs2\times0\farcs09$ \\
HD224151 & O54308010 & 02-18-99& 1496 &  $0\farcs1\times0\farcs03$ \\
HDE303308 & O4QXO4010 & 03-19-98& 2220 & $0\farcs2\times0\farcs09$ \\
          & O4QXO4020 & 03-19-98& 1560 & $0\farcs2\times0\farcs09$ \\
\enddata
\tablenotetext{a}{All the {\it HST} observations have been made through the STIS FUV-MAMA E140H grating.}
\end{deluxetable}

\begin{deluxetable}{lcccccc}
\tablecolumns{4}
\tablewidth{0pc}
\tablecaption{Summary of {\it FUSE} observations.}
\tablehead{ 
\colhead{Name} & \colhead{Program ID} & \colhead{Date} & \colhead{Aperture} & \colhead{Exp. time (Ksec)} & \colhead{\# exp.} & \colhead{Mode\tablenotemark{a}}\\
}
\startdata
HD75309  & P1022701 & 01-26-00  & LWRS  &  4.7  & 8  &  HIST \\
HD88115  & P1012301 & 04-04-00  & LWRS  &  4.5  & 8  &  HIST \\
HD91824  & A1180802 & 06-02-00  & LWRS  &  4.6  & 6  &  HIST \\
HD93205  & P1023601 & 02-01-00  & LWRS  &  4.7  & 7  &  HIST \\
HD93222  & P1023701 & 02-03-00  & LWRS  &  3.9  & 4  &  HIST \\
HD94493  & P1024101 & 03-26-00  & LWRS  &  4.4  & 7  &  HIST \\
HD99857  & P1024501 & 02-05-00  & LWRS  &  4.3  & 7  &  HIST \\
HD104705 & P1025701 & 02-05-00  & LWRS  &  4.5  & 6  &  HIST \\
HD124314 & P1026201 & 03-22-00  & LWRS  &  4.4  & 6  &  HIST \\
HD157857 & P1027501 & 09-02-00  & LWRS  &  4.0  & 8  &  HIST \\
HD177989 & P1017101 & 08-28-00  & LWRS  &  10.3  & 20 &  HIST \\
HD185418 & P1162301 & 08-10-00  & LWRS  &  4.4  & 3  &  TTAG \\
HD192639 & P1162401 & 06-12-00  & LWRS  &  4.8  & 2  &  TTAG \\
HD202347 & P1028901 & 06-20-00  & LWRS  &  0.1  & 1  &  HIST \\
HD210809 & P1223101 & 08-05-00  & LWRS  &  5.5  & 10 &  HIST \\
HD210839 & P1163101 & 07-22-00  & LWRS  &  6.1  & 10 &  HIST \\
HD218915 & P1018801 & 07-23-00  & LWRS  &  5.4  & 10 &  HIST \\
HD224151 & P1224101 & 08-11-00  & LWRS  &  6.0  & 12 &  HIST \\
HD303308 & P1222601 & 05-25-00  & LWRS  &  6.1  & 9  &  HIST \\
         & P1222602 & 05-27-00  & LWRS  &  7.7  & 12 &  HIST \\
\enddata
\tablenotetext{a}{HIST stands for histogram mode and TTAG for time-tagged mode.}
\end{deluxetable}

\clearpage 

\begin{table}
\begin{footnotesize}
\begin{center}
\caption{STIS and {\it FUSE} measurements of hydrogen}
\begin{tabular}{cccccc}
\tableline\tableline
name& N(H I) & N(H$_2$) & N(H$_{total}$) & f(H$_{2}$) & T$_{01}$\\
& $10^{20}$ $cm^{-2}$ & $10^{20}$ $cm^{-2}$ & $10^{20}$ $cm^{-2}$ & & (K)\\
\tableline
HD75309 &11.2 (2.0) & 1.5 (0.2) & 14.2 (2.0) & 0.21 & 65\\
HD88115 &10.0 (2.0)& 0.2 (0.1) & 10.3 (2.0) & 0.03 & 145 \\
HD91824 &11.7 (1.7) & 0.7 (0.2) & 13.1 (1.7) & 0.11 & 65 \\
HD93205 &24.0 (2.3)& 0.6 (0.1) & 25.3 (2.3) & 0.05 & 75\\
HD93222 &25.1 (3.7)& 0.6 (0.1) & 26.4 (3.7) & 0.05 & 121 \\
HD94493 &  12.0 (1.2)& 1.4 (0.2) & 14.8 (1.3) & 0.19 & 53\\
HD99857 & 17.4 (3.0)& 2.7 (0.4) & 22.8 (3.1) & 0.24 & 53\\
HD104705& 12.6 (2.5)& 1.2 (0.1) & 15.0 (2.6) & 0.16 & 137\\
HD124314&25.7 (5.2)& 3.3 (0.4) & 32.3 (5.3) & 0.20 & 54\\
HD157857& 18.6 (2.3)& 4.5 (0.6)& 27.6 (2.6) & 0.33 & 49\\
HD177989&  9.1 (1.8)& 1.5 (0.2)& 12.1 (1.8) & 0.25 & 88\\
HD185418& 14.1 (2.5)& 5.1 (1.3)& 24.3 (3.6) & 0.42 & 101 \\
HD192639& 19.5 (3.4)& 5.5 (1.5)& 30.5 (4.5) & 0.36 & 98 \\
HD202347&  8.7 (1.8)& 0.9 (0.2)& 10.5 (1.8) & 0.17 & 116\\
HD210809& 17.8 (4.1)& 1.3 (0.3)& 20.4 (4.1) & 0.13 & 166\\
HD210839& 15.5 (1.9)& 6.5 (1.0)& 28.5 (2.8) & 0.46 & 92\\
HD218915& 14.8 (1.4)& 1.6 (0.2)& 18.0 (1.5) & 0.18 & 56\\
HD224151& 20.9 (3.6)& 4.1 (0.6)& 29.1 (3.8) & 0.28 & 42\\
HD303308& 25.7 (4.5)& 2.2 (0.4)& 30.1 (4.6) & 0.15 & 54\\
\tableline
\tablecomments{The H~{\small I} column
density has been analysed using the STIS data and the continuum reconstruction method
while the total H$_2$ is given by profile fitting in the {\it FUSE}~range. Note that the stellar
atmospheres have small H~{\small I} column densities compared to the ISM. Toward HD210809,
the high H$_2$ rotational temperature shows that this sight line is mainly composed
of diffuse molecular components. The quoted errors are 1~$\sigma$.}
\end{tabular}
\end{center}
\end{footnotesize}
\end{table}

\begin{table}
\begin{footnotesize}
\begin{center}
\caption{Comparison of our measurements with previous measurements.}
\begin{tabular}{llccc}
\tableline\tableline
Species & Stars & log N & log N & Reference \\
 & & this work & literature & \\
\tableline
O~{\small I} & HD104705 & 17.81$^{+0.02}_{-0.03}$ &17.83$\pm~0.03$ & Howk et al. 2000b \\
`` &HD177989 & 17.79$^{+0.02}_{-0.03}$ &17.84$\pm~0.03$ & `` \\
`` &HD218915 & 17.82$^{+0.03}_{-0.03}$ &17.97$\pm~0.03$ & `` \\
`` &HD303308 & 18.09$^{+0.02}_{-0.03}$&18.10$\pm~0.04$ & `` \\
`` & HD192639 & 18.13$^{+0.03}_{-0.02}$&18.16$^{+0.11}_{-0.11}$ & Sonnentrucker et al. 2002 \\
`` & HD75309 & 17.72$^{+0.03}_{-0.04}$& 17.73$^{+0.05}_{-0.06}$ & Cartledge et al. 2001 \\
`` & HD185418 & 18.05$^{+0.02}_{-0.02}$&18.07$^{+0.05}_{-0.06}$ & `` \\
Kr~{\small I}& HD75309  & 12.24$^{+0.06}_{-0.08}$ &12.21$^{+0.09}_{-0.12}$& `` \\
`` & HD185418 & 12.48$^{+0.03}_{-0.04}$&12.50$^{+0.06}_{-0.07}$ & `` \\
H$_2$ & HD185418 &  20.71$^{+0.10}_{-0.13}$ & 20.76 $\pm~0.05$& Rachford et al. 2002 \\
`` & HD192639 &  20.74$^{+0.11}_{-0.14}$ &20.69 $\pm$~0.05 & `` \\
H~{\small I}& HD88115  &        21.00$^{+0.08}_{-0.08}$ & 21.01 $\pm$~0.11 & Diplas \& Savage 1994 \\
`` & HD93205  & 21.38$^{+0.04}_{-0.04}$ & 21.33 $\pm$~0.10 & ``\\
`` & HD94493  & 21.08$^{+0.04}_{-0.04}$ & 21.11 $\pm$~0.09 & ``\\
`` & HD99857  & 21.24$^{+0.05}_{-0.07}$ & 21.31 $\pm$~0.12 & ``\\
`` & HD104705 & 21.10$^{+0.06}_{-0.08}$& 21.11 $\pm$~0.07 & ``\\
`` & HD124314 & 21.41$^{+0.05}_{-0.08}$& 21.34 $\pm$~0.10 & ``\\
`` & HD157857 & 21.27$^{+0.05}_{-0.05}$ & 21.30 $\pm$~0.09 & ``\\
`` & HD177989 & 20.96$^{+0.06}_{-0.08}$& 20.95 $\pm$~0.09 & ``\\
`` & HD192639 & 21.29$^{+0.06}_{-0.07}$& 21.32 $\pm$~0.12 & ``\\
`` & HD210809 & 21.25$^{+0.06}_{-0.09}$& 21.25 $\pm$~0.07 & ``\\
`` & HD210839 & 21.19$^{+0.05}_{-0.04}$ & 21.15 $\pm$~0.12 & ``\\
`` & HD218915 & 21.17$^{+0.03}_{-0.05}$& 21.11 $\pm$~0.13 & ``\\
`` & HD224151 & 21.32$^{+0.05}_{-0.07}$& 21.32 $\pm$~0.10 & ``\\
`` & HD303308 & 21.41$^{+0.05}_{-0.07}$ & 21.45 $\pm$~0.09 & ``\\
\tableline
\tablecomments{The table shows that there is an excellent
agreement between our results and most of the previously published
values. For O~{\small I} in HD218915 the difference between our value and
the Howk et al. (2000b) value arises from the saturation correction applied in their
paper (see text). Quoted errors are 1~$\sigma$.}
\end{tabular}
\end{center}
\end{footnotesize}
\end{table}

\clearpage 

\begin{table}
\begin{footnotesize}
\begin{center}
\caption{STIS gas-phase measurements of oxygen.}
\begin{tabular}{ccccccc}
\tableline\tableline
name& W$_{\lambda}$ (1355) & O~{\small I} (AOD)& O~{\small I} (fit)& O~{\small I} (adopted)& ${\rm O/H}_{gas}$ & ${\rm O/H}_{dust}$\\
 & m \AA &  $10^{17}$ $cm^{-2}$ & $10^{17}$ $cm^{-2}$  & $10^{17}$ $cm^{-2}$ & ppm & ppm\\
\tableline
HD75309 &  9.0 (0.7)& 5.3 (0.4) & 5.2 (0.3)  & 5.2 (0.4)& 366 (60) & 151\\
HD88115 &  8.8 (1.1)& 5.1 (0.6) & 4.4 (0.6) & 4.7 (0.6) & 454 (117)& 62 \\
HD91824 &  11.0(1.4)& 6.2 (0.7) & 6.5 (0.4)  & 6.4 (0.7)& 488 (84) & 29 \\
HD93205 & 19.5 (1.1)& 8.5 (0.6) & 9.4 (0.5)  & 9.0 (0.6)& 356 (40) & 161\\
HD93222 & 25.8 (1.2)& 13.0 (0.6)& 14.0 (0.4) & 13.5 (0.6)& 512 (75)& 5 \\
HD94493 &  12.2(0.7)& 6.6 (0.6) & 7.5 (0.6)  & 7.0 (0.6)& 473 (57) & 44\\
HD99857 & 14.1 (0.8)& 8.0 (0.5) & 7.6 (0.5)  & 7.8 (0.5)& 342 (51) & 175\\
HD104705& 11.2 (0.7)& 6.4 (0.3) & 6.3 (0.3)  & 6.4 (0.3)& 427 (74) & 90\\
HD124314& 25.3 (1.0)& 15.3 (0.6)& 15.4 (0.8) & 15.3 (0.8)& 474 (81)& 43 \\
HD157857& 17.7 (0.8)& 11.6 (0.4)& 12.1 (0.4) & 11.9 (0.4)& 431 (43)& 86 \\
HD177989& 10.5 (0.6)& 6.1 (0.3) & 6.0 (0.4)  & 6.1 (0.4)& 504 (84) & 13\\
HD185418& 18.0 (0.7)& 11.4 (0.4)& 11.0 (0.3) & 11.2 (0.4)& 461 (70)& 56 \\
HD192639& 22.1 (0.9)& 13.5 (0.3)& 13.5 (0.6) & 13.5 (0.6)& 443 (69)& 74 \\
HD202347&  6.8 (1.1)& 3.8 (0.6) & 3.8 (0.4)  & 3.8 (0.6) & 362 (85)& 155 \\
HD210809& 12.5 (1.4)& 7.1 (0.7) & 5.6 (0.6)  & 6.3 (0.7) & 309 (71)& 208 \\
HD210839& 21.6 (0.9)& 13.4 (0.5)& 12.3 (0.6) & 13.0 (0.6)& 456 (49)&61 \\
HD218915& 12.0 (0.8)& 6.8 (0.4) & 6.5 (0.2)  & 6.6 (0.4)& 367 (37) &150\\
HD224151& 18.5 (0.9)& 12.2 (0.8)& 10.7 (0.7) &11.4 (0.8) & 392 (58)& 125\\
HD303308& 21.5 (1.1)& 12.7 (0.6)& 11.8 (0.4) &12.2 (0.6)& 405 (65) &112\\
\tableline
\tablecomments{The final column density
adopted for oxygen is a weighted average between AOD and profile fitting. The quoted errors
are 1~$\sigma$. The error on the ${\rm O/H}_{dust}$ estimates are not quoted since there could
be additional systematic errors in ${\rm O/H}_\odot$ due to the use of the average of
Allende et al. (2001) and Holweger (2001).}
\end{tabular}
\end{center}
\end{footnotesize}
\end{table}

\clearpage 

\begin{table}
\begin{footnotesize}
\begin{center}
\caption{STIS gas-phase measurements of krypton.}
\begin{tabular}{ccc}
\tableline\tableline
Stars & Kr~{\small I} & log O/Kr \\
 & 10$^{11}$ $cm^{-2}$ &\\
\tableline
HD75309 & 17.4 (2.8)& 5.47$^{+0.07}_{-0.08}$\\
HD99857 & 19.6 (4.1)& 5.62$^{+0.09}_{-0.11}$\\
HD104705\tablenotemark{*}& 13.3 (4.9)& 5.50$^{+0.14}_{-0.20}$\\
HD185418& 30.1 (2.5)& 5.57$^{+0.04}_{-0.05}$\\
\tableline
\tablenotetext{*}{In the case of HD104705, we actually
calculated the ratio in the main component (-0.3 km~s$^{-1}$) for which the krypton counterpart is detected }
\tablecomments{The analysis of the krypton $\lambda$1235 follows that of oxygen,
combining profile fitting and AOD. The quoted error bars are 1~$\sigma$. These results compare well
with the GHRS mean, log O/Kr = 5.56 $\pm$ 0.04 (Cartledge et al. 2001).}
\end{tabular}
\end{center}
\end{footnotesize}
\end{table}

\clearpage 

\begin{table}
\begin{footnotesize}
\begin{center}
\caption{Statistical analysis of the three samples.}
\begin{tabular}{lccccc}
\tableline\tableline
Sample & \# & $<$ O/H $>$ & Error in the mean & Standard deviation & $\overline{\chi^2}$ \\
 & & & &\\
\tableline
A         & 19 & 408 & 13 &59 & 0.9\\
B         & 32 & 377 & 10 &65 & 1.1\\
C         & 37 & 362 & 11 &73 & 1.4\\
\tableline
\tablecomments{ Sample A is made of our 19 sight lines. Sample B consists
of our 19 sight lines plus 13 sight lines within 1 kpc from Meyer et al. (1998).
Sample C is the same as sample B plus the 5 sight lines showing enhanced
depletion from Cartledge et al.(2001): 37 stars.}
\end{tabular}
\end{center}
\end{footnotesize}
\end{table}

\end{document}